\documentclass[a4paper,UKenglish]{lipics_md}
\usepackage{microtype}
\usepackage{graphicx}
\usepackage{tikz}
\usepackage{color}
\usepackage{ifpdf}

\title{The Weighted Gaussian Curvature Derivative
       of a Space-Filling Diagram\footnote{
        This project has received funding from the European Research Council (ERC)
        under the European Union's Horizon 2020 research and innovation programme
        (grant agreement No 78818 Alpha).
	It is also partially supported by the DFG Collaborative Research Center TRR 109,
	`Discretization in Geometry and Dynamics',
	through grant no.\ I02979-N35 of the Austrian Science Fund (FWF).}}

\author[1]{Arsenyi Akopyan}
\author[1]{Herbert Edelsbrunner}
\affil[1]{IST Austria (Institute of Science and Technology Austria),
	Klosterneuburg, \\
        Austria, \texttt{edels@ist.ac.at}, \texttt{akopjan@gmail.com}}
\authorrunning{A. Akopyan and H. Edelsbrunner}


\newcommand {\mm}[1] {\ifmmode{#1}\else{\mbox{\(#1\)}}\fi}

\newcommand {\scalprod}[2] {{\langle #1 , #2 \rangle}}
\newcommand{\denselist}{\itemsep 0pt\parsep=1pt\partopsep 0pt}

\newcommand{\ourproof}{\begin{proof}}
\newcommand{\eop}{\end{proof}}

\newcommand{\Bspace}       {\mm{{\mathbb B}}}
\newcommand{\Fspace}       {\mm{{\mathbb F}}}
\newcommand{\Mspace}[1]    {\mm{{\mathbb M}_{\rm {#1}}}}
\newcommand{\Rspace}       {\mm{{\mathbb R}}}
\newcommand{\Sspace}       {\mm{{\mathbb S}}}

\newcommand{\weight}[1]    {\mm{{w_{#1}}}}
\newcommand{\Vdom}[1]      {\mm{{V}_{#1}}}
\newcommand{\Gauss}        {\mm{\it Gauss}}
\newcommand{\Volume}       {\mm{\it Volume}}
\newcommand{\Area}         {\mm{\it Area}}
\newcommand{\Length}       {\mm{\it Length}}

\newcommand{\Card}[1]      {\mm{\#}{({#1})}}
\newcommand{\sgm}[1]       {\mm{\rm sgm}_{#1}}

\newcommand{\volume}       {\mm{\it vol}}
\newcommand{\area}         {\mm{\it area}}

\newcommand{\mean}         {\mm{\it mean}}
\newcommand{\gauss}        {\mm{\it gauss}}

\newcommand{\Euler}[1]     {\mm{\chi}{[{#1}]}}

\newcommand{\interior}[1]  {\mm{\rm int\,}{#1}}
\newcommand{\boundary}[1]  {\mm{\rm bd\,}{#1}}
\newcommand{\diff}         {\mm{\rm d}}
\newcommand{\intdiff}      {\mm{\rm \,d}}
\newcommand{\Diff}         {\mm{\rm D}}

\newcommand{\Edist}[2]     {\mm{\|{#1}-{#2}\|}}
\newcommand{\Fdist}[2]     {\mm{\|{#1}\!-\!{#2}\|}}

\newcommand{\Aalpha}[1]    {\mm{\alpha}^{#1}}
\newcommand{\aaa}          {\mm{\bf a}}
\newcommand{\bbb}          {\mm{\bf b}}
\newcommand{\ccc}          {\mm{\bf c}}
\newcommand{\ddd}          {\mm{\bf d}}
\newcommand{\eee}          {\mm{\bf e}}
\newcommand{\fff}          {\mm{\bf f}}
\newcommand{\GGG}          {\mm{\bf g}}
\newcommand{\hhh}          {\mm{\bf h}}
\newcommand{\mmm}          {\mm{\bf m}}
\newcommand{\nnn}          {\mm{\bf n}}

\newcommand{\ttt}          {\mm{\bf t}}

\newcommand{\uuu}          {\mm{\bf u}}
\newcommand{\vvv}          {\mm{\bf v}}
\newcommand{\xxx}          {\mm{\bf x}}
\newcommand{\zzz}          {\mm{\bf z}}
\newcommand{\TTT}          {\mm{\bf T}}

\newcommand{\ourparagraph}[1] {\vspace{0.1in} \noindent \textbf{#1}}
\newcommand{\Skip}[1]      {}

\begin{document}
\maketitle

\begin{abstract}
  The morphometric approach \cite{HRC13,RHK06} writes the solvation free energy
  as a linear combination of weighted versions of the volume, area,
  mean curvature, and Gaussian curvature of the space-filling diagram.
  We give a formula for the derivative of the weighted Gaussian curvature.
  Together with the derivatives of the weighted volume in \cite{EdKo03}, the
  weighted area in \cite{BEKL04},
  and the weighted mean curvature in \cite{AkEd19},
  this yields the derivative of the morphometric expression of
  solvation free energy.
\end{abstract}

\section{Introduction}
\label{sec:1}

This is a companion paper to \cite{EdKo03} and \cite{BEKL04} on the
derivatives of the weighted volume and of the weighed area
of a space-filling diagram,
and particularly of \cite{AkEd19} on the derivative of the weighted
mean curvature, with which this paper shares the notation.
Together with these three papers, we complete the explicit description
of the derivatives of the weighted intrinsic volumes in $\Rspace^3$.
All four derivatives are needed to realize the morphometric approach
of Mecke and Roth
to the \emph{solvation free energy}; see \cite{HRMD07,HRC13,KRM04,Mec96,RHK06}.
Getting its inspiration from the theory of intrinsic volumes,
this approach writes the solvation free energy as a linear combination
of the weighted volume, the weighted area, the weighted mean curvature,
and the weighted Gaussian curvature.
Compare this with Hadwiger's characterization theorem,
which asserts that any measure of convex bodies in $\Rspace^3$
that is invariant under rigid motion, continuous, and additive
is a linear combination of the four (unweighted) intrinsic volumes~\cite{Had51}.

Among the four intrinsic volumes, the Gaussian curvature is distinguished
by the Gauss--Bonnet theorem, which asserts that it is $2 \pi$ times
the Euler characteristic of the surface \cite{Bon48}.
Translated to a set of spheres in motion,
this implies that the curvature function is piecewise constant
and thus lacks any interesting derivative.
The situation changes drastically in the weighted case,
in which the contribution of a sphere to the Gaussian curvature
is scaled by a real weight that represents a physical property
of the corresponding atom, such as its hydrophobicity.
The main result of this paper is an explicit formula for the
derivative of the weighted Gaussian curvature of a space-filling diagram,
and an analysis of the cases in which this derivative either
does not exist or is not continuous.
In the special case in which all weights are equal, we still get
zero derivative almost everywhere and jumps with at places of
undefined derivative when the space-filling diagram undergoes
topological changes.

\ourparagraph{Outline.}
Section \ref{sec:2} reviews the background relevant to this paper.
Section \ref{sec:3} discusses the weighted Gaussian curvature of a space-filling diagram.
Section \ref{sec:4} derives the constituents of the weighted Gaussian curvature.
Section \ref{sec:5} states the derivative in terms of the gradient.
Section \ref{sec:6} analyzes the configurations at which the gradient
is not continuous.
Section \ref{sec:7} concludes this paper.

\section{Background}
\label{sec:2}

This section reviews the space-filling diagram and its dual alpha shape,
both geometric models of molecules.

\begin{figure}[hbt]
  \centering \resizebox{!}{2.2in}{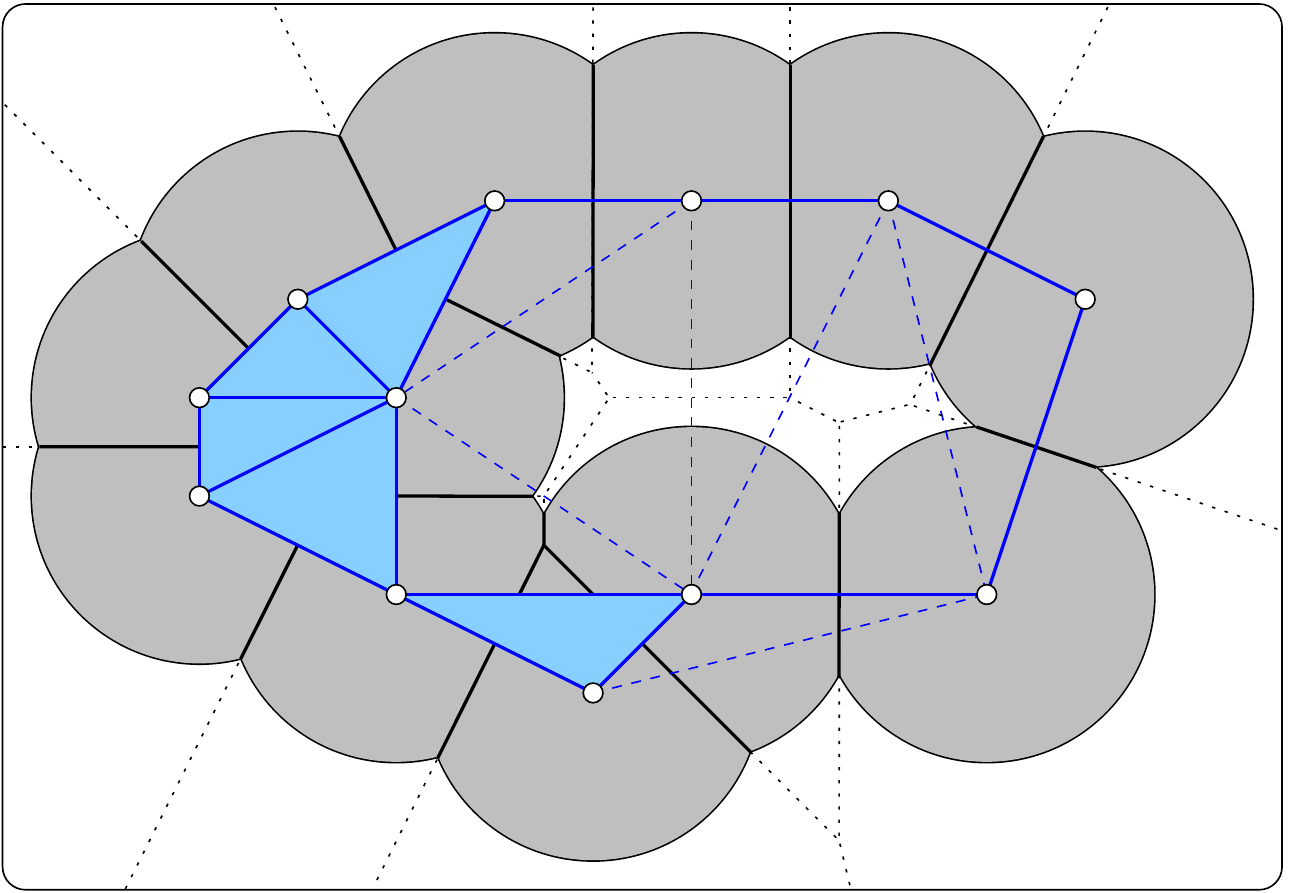}
  \caption{The Voronoi tessellation and Delaunay mosaic of
    a small number of point in the plane superimposed on each other.
    The part of the tessellation decomposing the union of disks into
    convex pieces is emphasized, and similarly the Alpha complex,
    which is the subcomplex of the Delaunay mosaic dual to this
    decomposition.}
  \label{fig:Alpha}
\end{figure}

\ourparagraph{Space-filling diagrams.}
Throughout this paper, we let $X$ be a collection of $n$ closed balls
in $\Rspace^3$,
and to remind us of the connection to chemistry and biology,
we call the union of these balls a \emph{space-filling diagram}.
Writing $x_i \in \Rspace^3$ for the center and $r_i > 0$ for the radius,
the corresponding ball is
$B_i = \{ a \in \Rspace^3 \mid \Edist{a}{x_i} \leq r_i \}$,
and the space-filling diagram is $\bigcup X = \bigcup_{i=0}^{n-1} B_i$;
see the $2$-dimensional case illustrated in Figure \ref{fig:Alpha},
in which $\bigcup X$ is a union of disks.
Its boundary consists of \emph{(spherical) patches} of the form
$S_i \setminus \bigcup_{j \neq i} \interior{B_j}$,
in which $S_i = \boundary{B_i}$.
These patches meet along closed \emph{(circular) arcs} of the form
$S_{ij} \setminus \bigcup_{k \neq i, j} \interior{B_k}$,
in which $S_{ij} = S_i \cap S_j$.
Finally, these arcs meet at \emph{corners} of the form
$S_{ijk} \setminus \bigcup_{\ell \neq i,j,k} \interior{B_\ell}$,
in which $S_{ijk} = S_i \cap S_j \cap S_k$.
For ease of reference, we collect and extend the introduced notation
in Table \ref{tbl:Notation}, which is given in the appendix.

We are interested in the volume of the space-filling diagram
and the area, mean curvature, and Gaussian curvature of its boundary.
Since this boundary is not necessarily smoothly embedded in $\Rspace^3$,
we use discrete formulas for the curvatures that agree with the
smooth formulas in the limit.
For the mean curvature, we have non-zero contributions from the
patches and the arcs; see Equation \eqref{eqn:mean},
and for the Gaussian curvature, we have non-zero contributions from
the patches, the arcs, and the corners; see Equation \eqref{eqn:gauss}.
However, according to the Gauss--Bonnet theorem \cite{Bon48},
the Gaussian curvature is invariant under deformations that
preserve its topology, and it changes by integer multiples of $2 \pi$
whenever the topology is not preserved.
The derivative under motion of the balls is therefore not very
informative, namely generically zero and occasionally undefined.
This is in sharp contrast of the weighted case.

\ourparagraph{Voronoi decomposition and dual alpha complex.}
We need a decomposition of the space-filling diagram and of its
boundary to define what we mean by its weighted measures.
Recall that $B_i$ is the closed ball with center $x_i$ and radius $r_i$.
The \emph{power distance} of~a~point~$a \in \Rspace^3$
from $B_i$ is $\pi_i (a) = \Edist{a}{x_i}^2 - r_i^2$.
Note that $B_i = \pi_i^{-1} (- \infty, 0]$.
The \emph{Voronoi domain} of $B_i$ is the set $V_i$ of points $a \in \Rspace^3$
for which $\pi_i (a) \leq \pi_j (a)$ for $0 \leq j < n$.
It is a possibly unbounded and always closed and convex
$3$-dimensional polyhedron;
see Figure \ref{fig:Alpha} for the $2$-dimensional case in which
the Voronoi domains are convex polygons.
Generically, $V_{ij} = V_i \cap V_j$ is either empty or $2$-dimensional,
$V_{ijk} = V_i \cap V_j \cap V_k$ is either empty or $1$-dimensional,
and $V_{ijk\ell} = V_i \cap V_j \cap V_k \cap V_\ell$ is either empty
or a point.
The Voronoi domains decompose the space-filling diagram into
convex sets of the form $B_i \cap V_i$.
The boundary of such a set consist of a spherical patch (which also
belong to the boundary of the space-filling diagram) and a~collection
of flat patches (along which the sets meet).
The spherical patch meets the flat patches along circular arcs
on the boundary of $\bigcup X$, and the flat patches meet along straight
line segments inside $\bigcup X$.

We use the notation in Table \ref{tbl:Notation} for the fractional
measures of the pieces in the decomposition.
They are conveniently computed using the \emph{alpha complex} of $X$,
which is the nerve of the collection of $B_i \cap V_i$ \cite{EdMu94}.
We recall that the Delaunay mosaic is the nerve of the collection
of $V_i$, which implies that the alpha complex is a subcomplex
of the Delaunay mosaic.
Just like the Delaunay mosaic, the alpha complex is geometrically realized
by mapping its vertices to the corresponding points in $\Rspace^3$.
The \emph{alpha shape} is the underlying space of the alpha complex,
and we stress that this is a set of points while the alpha complex is a
set of simplices.
The \emph{boundary} of the alpha shape consists of all points for which no
open neighborhood is contained in the set.
It is the underlying space of a subcomplex of the alpha complex,
and we refer to its members as \emph{boundary simplices}.
The alpha complex contains a simplex for every non-empty
common intersection of Voronoi domains, which makes it a convenient
book-keeping device for the intended computations.
To appreciate this facility, it is useful to familiarize ourselves with
the details of the correspondence between the boundaries of the
space-filling diagram and the alpha shape.
Whenever $S_i$ contributes a non-zero number of patches,
its center, $x_i$, is a boundary vertex of the alpha shape.
The number and structure of the patches can be extracted
from the structure of simplices in the alpha complex that share $x_i$.
Whenever $S_{ij}$ contributes a non-zero number of arcs,
the edge connecting $x_i$ to $x_j$ belongs to the boundary of the alpha shape.
The triangles and tetrahedra that share the edge are arranged cyclically
around the edge, and the number of gaps between contiguous triangles
in the cyclic order is equal to the number of contributed arcs.
Finally, whenever~$S_{ijk}$ contributes a non-zero number of
corners---namely one or two---the triangle
with vertices $x_i, x_j, x_k$ belongs to the boundary of the alpha shape.
The number of contributed corners is equal to the number of sides
of the triangle that face the outside of the alpha shape,
which is again one or two.
Similar but simpler rules of correspondence between the boundary
of the union of disks and the dual alpha shape in two dimensions
can be seen in Figure \ref{fig:Alpha}.

\section{Weighted Gaussian Curvature}
\label{sec:3}

According to the Gauss--Bonnet theorem \cite{Bon48},
the (integrated) Gaussian curvature of a closed surface, $\Fspace$, that is smoothly
embedded in $\Rspace^3$ is a fixed multiple of the Euler characteristic
of the surface:
\begin{align}
  \Gauss (\Fspace)  &=  \int_{x \in \Fspace} \kappa_1(x) \kappa_2(x) \intdiff x
                     =  2 \pi \Euler{\Fspace} ,
  \label{eqn:GaussBonnet}
\end{align}
in which $\kappa_1(x)$ and $\kappa_2(x)$ are the two principal curvatures at $x$.
The boundary of a space-filling diagram is generally not smoothly embedded,
but the extension of the Gaussian curvature used in this paper
still satisfies \eqref{eqn:GaussBonnet}.
This implies that the (unweighted) Gaussian curvature is constant
while the spheres move, until a time at which the surface changes topologically
and the Gaussian curvature jumps to a possibly different value.
In other words, the derivative of the unweighted Gaussian curvature
is generally zero, and undefined whenever the topology changes.
The weighted case is radically different because the contributions
of the individual spheres to the
weighted Gaussian curvature do not cancel each others changes.
In contrast to the unweighted case, the computation of the derivative
of the weighted Gaussian curvature of a space-filling diagram
is a meaningful as well as challenging undertaking.

\ourparagraph{Weighted intrinsic volume.}
Since the boundary of a space-filling diagram is only piecewise smooth,
we compute its weighted Gaussian curvature separately for the
sphere patches, the circular arcs, and the corners.
To facilitate a comparison, it is convenient to give such formulas
for the weighted versions of all four intrinsic volumes in $\Rspace^3$:
the weighted volume, the weighted area, the weighted mean curvature,
and the weighted Gaussian curvature.
Recall that $\bigcup X$ is the union of the balls $B_i$, for $0 \leq i < n$.
Its \emph{state}, $\xxx \in \Rspace^{3n}$, is the concatenation of the center vectors.
In other words, for $1 \leq \ell \leq 3$,
the $(3i+\ell)$-th coordinate of $\xxx$ is the $\ell$-th coordinate of $x_i$.
Decomposing the space-filling diagram with the Voronoi tessellation,
we write $\weight{i} \in \Rspace$ for the weight of $B_i$ and
use the notation introduced in Table \ref{tbl:Notation}
to give formulas for the weighted intrinsic volumes.
\begin{proposition}[Weighted Intrinsic Volumes]
  \label{prop:IntrinsicVolumes}
  The weighted volume, surface area, mean curvature, and Gaussian curvature
  of the space-filling diagram, $\bigcup X$, are
  \begin{align}
    \volume (\xxx) &=  \tfrac{4 \pi}{3} \sum_i \weight{i} \nu_i r_i^3,
      \label{eqn:volume} \\
    \area (\xxx)   &=  4 \pi \sum_i \weight{i} \sigma_i r_i^2 ,       
      \label{eqn:area}   \\
    \mean (\xxx)   &=  4 \pi \sum_i \weight{i} \sigma_i r_i
                     -   \pi \sum_{i,j} \tfrac{\weight{i}+\weight{j}}{2}
                                               \sigma_{ij} \phi_{ij} r_{ij} ,     
      \label{eqn:mean}   \\
    \gauss (\xxx)  &=  4 \pi \sum_i \weight{i} \sigma_i 
                     -   \pi \sum_{i,j} \tfrac{\weight{i}+\weight{j}}{2}
                                               \sigma_{ij} \lambda_{ij}
                     + \tfrac{1}{3} \sum_{i,j,k}
                          (\alpha_i \weight{i} + \alpha_j \weight{j}
                                               + \alpha_k \weight{k})
                           \sigma_{ijk} \phi_{ijk} .
      \label{eqn:gauss}
  \end{align}
\end{proposition}
The weighted Gaussian curvature function decomposes into three sums,
which account for the sphere patches, the circular arcs, and the corners, respectively.
The sums are over ordered index sequences.
This implies that the same three indices appear six times,
which explains the coefficient of $\tfrac{1}{3}$ in front of this sum.
We will see that $\phi_{i,jk} = \alpha_i \phi_{ijk}$ is the area of a~spherical
quadrangle, so we can simplify the last sum.
Formulas for the derivatives of $\sigma_i$ and of $\sigma_{ij}$
can be found in \cite{AkEd19}.
We also need derivatives of $\lambda_{ij}$
(the combined length of the projection of two normals; see Figure \ref{fig:lambda}),
and of $\phi_{i,jk}$
(the area of a spherical quadrangle),
which will be computed in Sections \ref{sec:41} and \ref{sec:42}.
Similar to the state, $\xxx$, we write $\ttt \in \Rspace^{3n}$
for the \emph{momentum}, which is obtained by concatenating the
velocity vectors of the $n$ balls in the order of their indices.
We write $\lambda_{ij} (\tau)$ for $\lambda_{ij}$ at state $\xxx + \tau \ttt$
and $\lambda_{ij}'$ for the derivative of $\lambda_{ij}$ at $\tau = 0$,
and similarly for $\sigma_{ijk}$ and $\phi_{i,jk}$.
Derivatives with respect to parameters other than $\tau$
are explicitly stated as such.

\ourparagraph{How to split a corner.}
Every corner of the space-filling diagram has a non-zero contribution
to the Gaussian curvature,
and since the spheres have weights,
we need to decide how to distribute this contribution among the
generically three spheres that meet at the corner.
We do this in a way that is consistent with splitting the curvature
concentrated along an arc in equal halves,
which is motivated by the Appolonius diagram of the spheres;
see \cite{AkEd19}.

\begin{figure}[b!]
  \centering \includegraphics{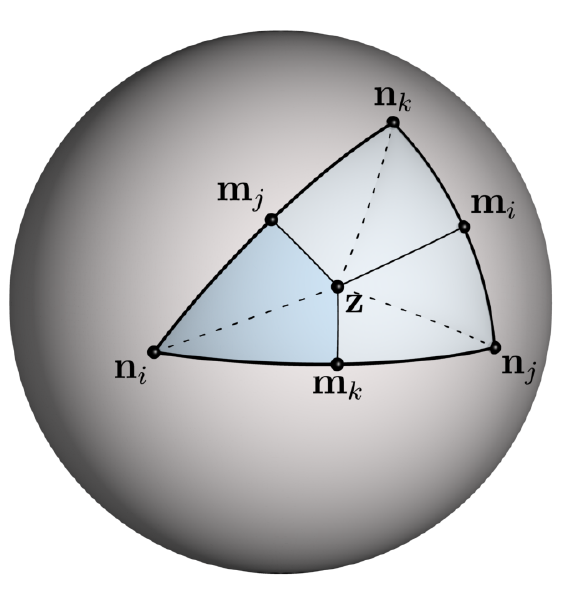}
  \caption{A spherical triangle with vertices $\nnn_i, \nnn_j, \nnn_k$
    and the cap with center $\zzz$ whose boundary is the unique circle that
    passes through the three vertices.
    The spherical triangle is decomposed into three spherical quadrangles
    by connecting $\zzz$ to the midpoints of the three sides,
    and into six cones by further connecting $\zzz$ to the three vertices.}
  \label{fig:split}
\end{figure}

Consider spheres $S_i, S_j, S_k$, and suppose that~$P$ is one of the two
points in which the spheres generically intersect.
Write $\nnn_i, \nnn_j, \nnn_k$ for the unit outward normals of the
spheres at~$P$, and recall that $\phi_{ijk}$ is the area of the
spherical triangle with vertices $\nnn_i, \nnn_j, \nnn_k$ on the unit sphere;
see Figure \ref{fig:split}.
We split the contribution of the corner using
$\alpha_i + \alpha_j + \alpha_k = 1$,
in which the three coefficients are the fractions of the areas of the
spherical quadrangles obtained by decomposing the spherical triangle
with the great-circle arcs that connect the center of the circumcircle,
denoted $\zzz$, with the midpoints of the sides, denoted $\mmm_i, \mmm_j, \mmm_k$.
By connecting $\zzz$ with the three vertices,
we further decompose each quadrangle into two spherical triangles,
which we refer to as \emph{cones} of $\zzz$ over the half-sides;
see again Figure \ref{fig:split}.
We remark that $\zzz$ can lie outside the spherical triangle,
in which case the areas of two of the cones are considered to be negative.
The cones come in pairs of equal area,
and each pair glues together to form an isosceles spherical triangle.
As remarked, exactly one of the three isosceles triangles has negative
area if $z$ is not contained in the original spherical triangle.
Since each of the three quadrangles is the union of two cones,
we can write its area as half the sum of the area of two
isosceles triangles:
\begin{align}
  \Area (\nnn_i \mmm_k \zzz \mmm_j)
    &=  \tfrac{1}{2} \left[ \Area (\nnn_i \nnn_j \zzz)
                          + \Area (\nnn_k \nnn_i \zzz) \right] ,
  \label{eqn:quadranglearea}
\end{align}
and similarly for $\Area (\nnn_j \mmm_i \zzz \mmm_k)$ and
$\Area (\nnn_k \mmm_j \zzz \mmm_i)$.
Equivalently, the area of the quadrangle is half of
$\Area (\nnn_i \nnn_j \nnn_k) - \Area (\nnn_j \nnn_k \zzz)$,
but we prefer \eqref{eqn:quadranglearea} for our computations.

\section{Derivatives}
\label{sec:4}

We reuse the expressions for $\sigma_i'$ and $\sigma_{ij}'$ in \cite{AkEd19},
and we develop formulas for $\lambda_{ij}'$, $\sigma_{ijk}'$, $\phi_{i,jk}'$.
The middle of the three, $\sigma_{ijk}'$, is piecewise zero because
$\sigma_{ijk}$ is piecewise constant.
As will be discussed in Section \ref{sec:6}, $\sigma_{ijk}$
changes at non-generic states, where its derivative is not defined.
We first consider the derivative of $\lambda_{ij}$ and second that
of $\phi_{i,jk}$.

\subsection{Derivative of $\lambda_{ij}$}
\label{sec:41}

\begin{figure}[hbt]
  \centering \resizebox{!}{2.0in}{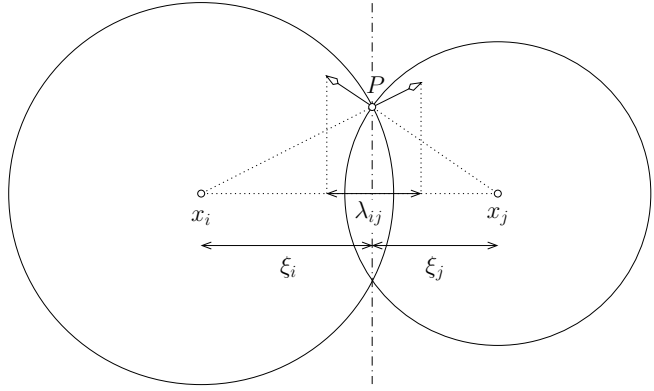}
  \caption{After projecting the unit normal vectors at an intersection point,
    $\lambda_{ij}$ is the distance between the two endpoints.}
  \label{fig:lambda}
\end{figure}

Given spheres $S_i$ and $S_j$, we define $\lambda_{ij}$ as the combined
length of the unit normal vectors at a point $P \in S_i \cap S_j$
after projection to the line that passes through the centers, $x_i$ and~$x_j$;
see Figure \ref{fig:lambda}.
If the two vectors point in the same direction---after projection---then
we take one length negative,
so more precisely, $\lambda_{ij}$ is the distance between the projections
of the endpoints of the unit vectors.
To compute the derivative of $\lambda_{ij}$, it will be useful to recall
that the signed distance of $x_i$ from the Voronoi plane of the two spheres is
\begin{align}
  \xi_i  &=  \frac{1}{2} \left( \Edist{x_i}{x_j}
                              + \frac{r_i^2 - r_j^2}{\Edist{x_i}{x_j}} \right) ;
\end{align}
see \cite{AkEd19}.
The length of the unit normal vector of $S_i$ at $P$, after projection,
is $\xi_i / r_i$, and since~$\xi_i$ is signed so is this length, as required.
Adding the two signed lengths, we get
\begin{align}
  \lambda_{ij}  &=  \frac{\xi_i}{r_i} + \frac{\xi_j}{r_j}
                 =  \frac{1}{2 r_i} \left( \Edist{x_i}{x_j} + \frac{r_i^2-r_j^2}{\Edist{x_i}{x_j}} \right)
                 +  \frac{1}{2 r_j} \left( \Edist{x_i}{x_j} - \frac{r_i^2-r_j^2}{\Edist{x_i}{x_j}} \right).
\end{align}
The derivative with respect to the distance between the two centers is therefore
\begin{align}
  \frac{\diff \lambda_{ij}}{\diff \Edist{x_i}{x_j}} 
    &=  \left( \frac{1}{2 r_i} + \frac{1}{2 r_j} \right)
      - \left( \frac{1}{2 r_i} - \frac{1}{2 r_j} \right) \frac{r_i^2-r_j^2}{\Edist{x_i}{x_j}^2} .
  \label{eqn:dlambda}
\end{align}
To translate this result to the derivative with respect to the motion
vectors $\ttt_i$ of~$x_i$ and $\ttt_j$ of~$x_j$,
we note that $\uuu_{ij} = (x_i-x_j) / \Edist{x_i}{x_j}$
is the unit vector that points from $x_j$ to $x_i$.
\begin{lemma}[Derivative of $\lambda_{ij}$]
  \label{lem:dlambdaij}
  The derivative of the combined length of the unit normal vectors at an
  intersection point of spheres $S_i$ and
  $S_j$---after projection to the line passing through the centers---at
  state $\xxx$ with momentum $\ttt$ is
  \begin{align}
    \lambda_{ij}'  &=  \frac{\diff \lambda_{ij}}{\diff \Edist{x_i}{x_j}}
                       \scalprod{\uuu_{ij}}{\ttt_i - \ttt_j} ,
  \end{align}
  in which the coefficient is given in \eqref{eqn:dlambda}.
\end{lemma}

\subsection{Derivative of $\phi_{i,jk}$}
\label{sec:42}

Instead of deriving $\alpha_i$ and $\phi_{ijk}$ separately,
we recall that
$\phi_{i,jk} = \alpha_i \phi_{ijk}$ is the area of a~spherical quadrangle,
which we can derive directly.
We begin with intrinsic formulas for the area and the radius of
a spherical triangle.

\ourparagraph{Area of spherical triangle.}
Recall that $\phi_{ijk}$ is the area and $\phi_{ij}, \phi_{jk}, \phi_{ki}$
are the lengths of the sides of the spherical triangle with vertices
$\nnn_i, \nnn_j, \nnn_k$.
Writing $s = \tfrac{1}{2} ( \phi_{ij} + \phi_{jk} + \phi_{ki} )$ for half
the perimeter, we have the following formula for the sine of half the area:
\begin{align}
  \sin \frac{\phi_{ijk}}{2}  &=  \frac{ \sqrt{ \sin s \sin (s-\phi_{ij})
                                                     \sin (s-\phi_{jk})
                                                     \sin (s-\phi_{ki}}) }
                                {2 \cos \frac{\phi_{ij}}{2}
                                   \cos \frac{\phi_{jk}}{2}
                                   \cos \frac{\phi_{ki}}{2}};
  \label{eqn:sphericalarea}
\end{align}
see \cite[page 230, equation (225)]{Cha87}.
To rewrite this formula, we introduce the squared cosines of the
half-lengths as parameters.
Geometrically, $\cos^2 \tfrac{x}{2}$ is the distance from the center
to the midpoint of the straight edge that connects the endpoints
of an arc of length $x$ on the unit sphere.
The square of this distance conveniently relates to Pythagoras' theorem
for right-angled triangles.
In a first step, we rewrite the product of the four sines.
\begin{formula}[Product of Sines]
  \label{form:ProductofSines}
  Writing $a = \cos^2 \frac{\phi_{ij}}{2}$,
  $b = \cos^2 \frac{\phi_{jk}}{2}$, $c = \cos^2 \frac{\phi_{ki}}{2}$,
  the expression under the square root in \eqref{eqn:sphericalarea} satisfies
  \begin{align}
    \sin s \sin (s-\phi_{ij}) \sin (s-\phi_{jk}) \sin (s-\phi_{ki})
      &=  4abc - (a+b+c-1)^2 .
    \label{eqn:formula5}
  \end{align}
\end{formula}
We refer to Appendix \ref{app:A} for a proof.
We can thus rewrite \eqref{eqn:sphericalarea} to give the area
of any spherical triangle given the squared cosines of its
half-lengths:
\begin{align}
  S(a,b,c)  &=  2 \arcsin \sqrt{ \frac{4abc - (a+b+c-1)^2}{4abc} } .
\end{align}
By \eqref{eqn:sphericalarea} and \eqref{eqn:formula5},
we have $\phi_{ijk} = S(a,b,c)$ with parameters as given in
Formula \ref{form:ProductofSines}.
We introduce notation for the areas of the isosceles triangles
obtained by gluing equal-area cones:
$A(a,r) = S(a,r,r)$, $B(b,r) = S(r,b,r)$, $C(c,r) = S(r,r,c)$,
with $r = \cos^2 \frac{R_{ijk}}{2}$,
in which $R_{ijk}$ is the radius of the cap bounded ~by the
circle that passes through $\nnn_i, \nnn_j, \nnn_k$.
By symmetry between the arguments of $S$,
all three functions are the same, but it is convenient to distinguish them
so we do not lose the connection to the sides of the spherical triangle.
Using \eqref{eqn:quadranglearea}, we get the areas of the spherical quadrangles:
\begin{align}
  \phi_{i,jk}  &=  \tfrac{1}{2}
          \left[ \sgm{k,ij} \cdot A(a,r) + \sgm{j,ki} \cdot C(c,r) \right] ,
    \label{eqn:areaQi} \\
  \phi_{j,ki}  &=  \tfrac{1}{2}
          \left[ \sgm{i,jk} \cdot B(b,r) + \sgm{k,ij} \cdot A(a,r) \right] ,
    \label{eqn:areaQj} \\
  \phi_{k,ij}  &=  \tfrac{1}{2}
          \left[ \sgm{j,ki} \cdot C(c,r) + \sgm{i,jk} \cdot B(b,r) \right] ,
    \label{eqn:areaQk}
\end{align}
in which the signs depend on whether $\zzz$ lies inside or outside
the triangle spanned by $\nnn_i, \nnn_j, \nnn_k$.
Specifically, $\sgm{i,jk} = +1$ if $\zzz$ and $\nnn_i$ lie on the same
side of the great-circle that passes through $\nnn_j$ and $\nnn_k$,
and $\sgm{i,jk} = -1$ if the two points lie on opposite sides
of this great-circle.
In the boundary case, when $\zzz$ lies on the great-circle,
$\nnn_i$, $\nnn_j$, $\nnn_k$ form a Euclidean right-angled triangle.
Pythagoras' theorem for its edges is equivalent to
$\sin^2 \frac{\phi_{ij}}{2} + \sin^2 \frac{\phi_{ki}}{2} = 
 \sin^2 \frac{\phi_{jk}}{2}$.
If $\zzz$ and $\nnn_i$ lie on the same side of the great-circle,
then the left-hand side is strictly larger than the right-hand side.
Substituting $\sin^2 x = 1 - \cos^2 x$, we get
\begin{align}
  \sgm{i,jk}  &=  \left\{ \begin{array}{rl}
    +1  &  \mbox{\rm if~} \cos^2 \frac{\phi_{ij}}{2}
           + \cos^2 \frac{\phi_{ki}}{2} \leq 1 + \cos^2 \frac{\phi_{jk}}{2}, \\
    -1  &  \mbox{\rm otherwise}.
                  \end{array} \right.
\end{align}
Since $S(a,b,c)$ is symmetric in its three arguments, it suffices
to compute the derivative with respect to one of them.
To this end, we recall a few basic rules of differentiation,
namely $\diff \arcsin x / \diff x = 1 / \sqrt{1 - x^2}$,
$\diff \sqrt{x} / \diff x = 1 / \sqrt{4x}$,
and $\diff (\tfrac{u}{v}) / \diff x
      = (v \diff u / \diff x - u \diff v / \diff x) / v^2$
for the ratio of two functions in $x$.
Setting $u(a,b,c) = 4abc - (a+b+c-1)^2$ and $v(a,b,c) = 4abc$,
we get $\diff u / \diff a = 4bc - 2(a+b+c-1)$ and $\diff v / \diff a = 4bc$.
Furthermore setting $w = \tfrac{u}{v}$ and $t = \sqrt{w}$,
we have $S(a,b,c) = 2 \arcsin t$ and therefore
\begin{align}
  \frac{\diff S}{\diff t}  &=  \frac{2}{\sqrt{1-t^2}}
    =  \frac{4 \sqrt{abc}}{a+b+c-1} ,                              \\
  \frac{\diff t}{\diff w}  &=  \frac{1}{2 \sqrt{w}}
    =  \frac{\sqrt{abc}}{\sqrt{4abc - (a+b+c-1)^2}} ,     \\
  \frac{\diff w}{\diff a}  &=  \frac{v \diff u / \diff a - u \diff v / \diff a}
                                    {v^2}
    =  \frac{ -8abc (a+b+c-1) + 4bc (a+b+c-1)^2 }
            {(4 abc)^2} .
\end{align}
Multiplying the three right-hand sides and simplifying the result,
we get the derivative of the area
with respect to the squared cosine of one half-length:
\begin{align}
  \frac{\diff S}{\diff a} &= \frac{-a+b+c-1}{a \sqrt{4abc-(a+b+c-1)^2} }.
    \label{eqn:diffSdiffa} 
\end{align}
Permuting the three arguments, we get symmetric expressions
for $\diff S / \diff b$ and for $\diff S / \diff c$.

\ourparagraph{Radius of spherical triangle.}
We need a formula for the radius of the cap, $R_{ijk}$,
in the same parametrization:
\begin{align}
  \tan R_{ijk}      &=  \frac{\tan \frac{\phi_{ij}}{2}
                              \tan \frac{\phi_{jk}}{2}
                              \tan \frac{\phi_{ki}}{2}}
                             {\sin \frac{\phi_{ijk}}{2}} 
                     =  \frac{ 2 \sin \frac{\phi_{ij}}{2}
                                 \sin \frac{\phi_{ij}}{2}
                                 \sin \frac{\phi_{ij}}{2} }
                             { \sqrt{ \sin s \sin (s-\phi_{ij})
                                             \sin (s-\phi_{jk})
                                             \sin (s-\phi_{ki}}) } ;
    \label{eqn:tanR}
\end{align}
see \cite[page 246, equation (307)]{Cha87}.
Writing this in terms of the squared cosines, we get
\begin{align}
  \tan R(a,b,c)  &=  \sqrt{ \frac{4(1-a)(1-b)(1-c)}
                                 {4abc-(a+b+c-1)^2} } .
  \label{eqn:Radius}
\end{align}
To turn this into a formula for the squared cosine of the half-radius,
we recall two basic trigonometric identities:
$\cos^2 \tfrac{x}{2}  =  \tfrac{1}{2} [ 1 + \cos x ]
    =  \tfrac{1}{2} [ 1 + 1 / \sqrt{\tan^2 x + 1} ]$.
Writing $r(a,b,c) = \cos^2 \frac{R(a,b,c)}{2}$
and using \eqref{eqn:Radius}, this gives
\begin{align}
  r(a,b,c)  &=  \tfrac{1}{2} \left[ 1 +
      \sqrt{ \frac{4abc - (a+b+c-1)^2} {4(1-a)(1-b)(1-c) + 4abc - (a+b+c-1)^2}}
    \; \right]
    \label{eqn:Radius1} \\
    &=  \tfrac{1}{2} + \tfrac{1}{2}
      \sqrt{ \frac{4abc - (a+b+c-1)^2}
                 {(a-1)^2 + (b-1)^2 + (c-1)^2 - (a-b)^2 - (a-c)^2 - (b-c)^2} } .
    \label{eqn:Radius2}
\end{align}
Since $r(a,b,c)$ is symmetric in the three arguments,
it suffices to compute the derivative with respect to one of them.
Referring to \eqref{eqn:Radius2}, we set
\begin{align}
  u(a,b,c)  &=  4abc - (a+b+c-1)^2 ,                                       \\
  v(a,b,c)  &=  (a-1)^2 + (b-1)^2 + (c-1)^2 - (a-b)^2 - (a-c)^2 - (b-c)^2, 
\end{align}
which give ${\diff u}/{\diff a} =  4bc - 2(a+b+c-1)$
and ${\diff v}/{\diff a} =  2(-a+b+c-1)$.
Observing that
$r(a,b,c) = \tfrac{1}{2} + \tfrac{1}{2} \sqrt{ u(a,b,c) / v(a,b,c) }$,
we see that its derivative with respect to the first argument is
$(v \diff u / \diff a- u \diff v / \diff a)
   / \left( 4 v \sqrt{v} \sqrt{u} \right)$,
which implies
\begin{align}
  \frac{\diff r}{\diff a}
    &=  \frac{ (1-b)(1-c) \cdot [(a-1)^2-(b-c)^2] }
             { \sqrt{ u(a,b,c)} \sqrt{ v(a,b,c)}^3 } .
\end{align}
Symmetric expressions hold for $\diff r / \diff b$ and for $\diff r / \diff c$.

\ourparagraph{Using the chain rule.}
We get the derivatives of the areas of the spherical quadrangles by
combining the results with the chain rule.
Beyond the relations above, we need the derivative of the length
of a side, which we find in \cite[equation (30)]{AkEd19}:
\begin{align}
  \frac{\diff \phi_{ij}}{\diff \Edist{x_i}{x_j}}
    &=  \frac{2 \Edist{x_i}{x_j}}
             {\sqrt{ 2 (r_i^2+r_j^2) \Edist{x_i}{x_j}^2
                     - (r_i^2-r_j^2)^2 - \Edist{x_i}{x_j}^4 }} .
  \label{eqn:diffphiij}
\end{align}
We recall that
$\diff \cos^2 \frac{x}{2} / \diff x  = - \sin \frac{x}{2} \cos \frac{x}{2}$,
and we use \eqref{eqn:areaQi} to express the area of the spherical quadrangle:
$\phi_{i,jk} = \tfrac{1}{2} [ \sgm{k,ij} A(a,r) + \sgm{j,ki} C(c,r) ]$,
in which $a = \cos^2 \frac{\phi_{ij}}{2}$,
         $b = \cos^2 \frac{\phi_{jk}}{2}$,
         $c = \cos^2 \frac{\phi_{ki}}{2}$,
and      $r = \cos^2 \frac{R_{ijk}}{2}$ as given in \eqref{eqn:Radius2}.
The derivatives of $A$ with respect to its arguments are
$\diff A / \diff a = \diff S / \diff a$
and $\diff A / \diff r = \diff S / \diff b + \diff S /\diff c$,
both evaluated at $b = c = r$.
The derivatives with respect to the distances between the centers of the
spheres are therefore
\begin{align}
  \frac{\diff A}{\diff \Edist{x_i}{x_j}}
    &=  \frac{\diff A}{\diff a} \cdot \frac{\diff a}{\diff \phi_{ij}}
              \cdot \frac{\diff \phi_{ij}}{\diff \Edist{x_i}{x_j}}    
      + \frac{\diff A}{\diff r} \cdot
        \frac{\diff r}{\diff a} \cdot \frac{\diff a}{\diff \phi_{ij}}
              \cdot \frac{\diff \phi_{ij}}{\diff \Edist{x_i}{x_j}} ,
    \label{eqn:diffAdiffxixj} \\
  \frac{\diff A}{\diff \Edist{x_j}{x_k}}
    &=  \frac{\diff A}{\diff r} \cdot \frac{\diff r}{\diff b}
              \cdot \frac{\diff b}{\diff \phi_{jk}}
              \cdot \frac{\diff \phi_{jk}}{\diff \Edist{x_j}{x_k}},   
    \label{eqn:diffAdiffxjxk} \\
  \frac{\diff A}{\diff \Edist{x_k}{x_i}}
    &=  \frac{\diff A}{\diff r} \cdot \frac{\diff r}{\diff c}
              \cdot \frac{\diff c}{\diff \phi_{ki}}
              \cdot \frac{\diff \phi_{ki}}{\diff \Edist{x_k}{x_i}}.
    \label{eqn:diffAdiffxkxi}
\end{align}
\eqref{eqn:diffAdiffxjxk} and \eqref{eqn:diffAdiffxkxi} are symmetric
because the same two sides of the isosceles triangle depend
on $\Edist{x_j}{x_k}$ and on $\Edist{x_k}{x_i}$,
while \eqref{eqn:diffAdiffxixj} is different because
all three sides depend on $\Edist{x_i}{x_j}$.
We can now summarize and state the derivative of $\phi_{i,jk}$
with respect to the motion vector.
\begin{lemma}[Derivative of $\phi_{i,jk}$]
  \label{lem:dphiijk}
  The derivative of the area of the spherical quadrangle with vertices
  $\nnn_i, \mmm_k, \zzz, \mmm_j$ on the unit sphere is
  \begin{align}
    \phi_{i,jk}'  &=  p_{ij} \scalprod{\uuu_{ij}}{\ttt_i-\ttt_j}
                    + p_{jk} \scalprod{\uuu_{jk}}{\ttt_j-\ttt_k}
                    + p_{ki} \scalprod{\uuu_{ki}}{\ttt_k-\ttt_i} ,     \\
    p_{ij}        &=  \tfrac{1}{2}
      \left[ \sgm{k,ij} \cdot \frac{\diff A}{\diff \Edist{x_i}{x_j}}
           + \sgm{j,ki} \cdot \frac{\diff C}{\diff \Edist{x_i}{x_j}} \right], \\
    p_{jk}        &=  \tfrac{1}{2}
      \left[ \sgm{k,ij} \cdot \frac{\diff A}{\diff \Edist{x_j}{x_k}}
           + \sgm{j,ki} \cdot \frac{\diff C}{\diff \Edist{x_j}{x_k}} \right], \\
    p_{ki}        &=  \tfrac{1}{2}
      \left[ \sgm{k,ij} \cdot \frac{\diff A}{\diff \Edist{x_k}{x_i}}
           + \sgm{j,ki} \cdot \frac{\diff C}{\diff \Edist{x_k}{x_i}} \right],
  \end{align}
  in which the derivatives on the right-hand side are given in
  or symmetric to \eqref{eqn:diffAdiffxixj}, \eqref{eqn:diffAdiffxjxk},
  \eqref{eqn:diffAdiffxkxi}.
\end{lemma}
Starting with \eqref{eqn:areaQj}, \eqref{eqn:areaQk}
instead of \eqref{eqn:areaQi}, we get
\begin{align}
  \phi_{j,ki}' &= q_{ij} \scalprod{\uuu_{ij}}{\ttt_i-\ttt_j}
                + q_{jk} \scalprod{\uuu_{jk}}{\ttt_j-\ttt_k}
                + q_{ki} \scalprod{\uuu_{ki}}{\ttt_k-\ttt_i} ,  \\
  \phi_{k,ij}' &= s_{ij} \scalprod{\uuu_{ij}}{\ttt_i-\ttt_j}
                + s_{jk} \scalprod{\uuu_{jk}}{\ttt_j-\ttt_k}
                + s_{ki} \scalprod{\uuu_{ki}}{\ttt_k-\ttt_i} ,
\end{align}
with symmetrically defined coefficients.

\section{Gradients}
\label{sec:5}

For computational purposes, it is convenient to write the derivative
of the weighted Gaussian curvature function,
$\gauss \colon \Rspace^{3n} \to \Rspace$,
in terms of the gradient of $\gauss$ at $\xxx \in \Rspace^{3n}$,
denoted $\GGG = \nabla_{\!\xxx} \gauss$.
Recalling \eqref{eqn:gauss}, this derivative is
$\gauss' = d' + e' + f' + h'$, with
\begin{align}
  d'  &=  4 \pi \sum_i \weight{i} \sigma_i' ,
    \label{eqn:dd} \\
  e'  &=  \frac{\pi}{2} \sum_{i,j} (\weight{i}+\weight{j})
                           \sigma_{ij}' \lambda_{ij} , 
    \label{eqn:de} \\
  f'  &=  \frac{\pi}{2} \sum_{i,j} (\weight{i}+\weight{j})
                           \sigma_{ij} \lambda_{ij}' , 
    \label{eqn:df} \\
  h'  &=  \frac{\sigma_{ijk}}{3} \sum_{i,j,k}
          [ \weight{i} \phi_{i,jk}' + \weight{j} \phi_{j,ki}'
                                    + \weight{k} \phi_{k,ij}' ] .
    \label{eqn:dh}
\end{align}
Writing $\GGG = [g_1, g_2, \ldots, g_{3n}]^T$, we recall that
$\GGG_i = [g_{3i+1}, g_{3i+2}, g_{3i+3}]^T$ is the $3$-dimensional
gradient that applies to $x_i$.
Using boldface letters for the gradients of $d$, $e$, $f$, $h$,
and similar conventions for the $3$-dimensional sub-vectors,
we have $\GGG = \ddd + \eee + \fff + \hhh$ and
$\GGG_i = \ddd_i + \eee_i + \fff_i + \hhh_i$ for $0 \leq i < n$.
We get the gradients by redistributing the relevant derivatives
given in \cite{AkEd19} and stated in Lemmas \ref{lem:dlambdaij}
and \ref{lem:dphiijk}.

\ourparagraph{First term.}
The derivative of $\sigma_i$ is given in
\cite[equations (19) and (21)]{AkEd19} as
\begin{align}
  \sigma_i'  &=  \sum_j \frac{\sigma_{ij}}{4 r_i}
                 \left( 1 - \frac{r_i^2-r_j^2}{\Edist{x_i}{x_j}^2} \right)
                 \scalprod{\uuu_{ij}}{\ttt_i-\ttt_j}
               + \sum_{j,k} \frac{r_{ijk} \nu_{ijk}}
                                 {2 \pi r_i \Edist{x_i}{x_j}}
                 \scalprod{\uuu_{ijk}}{\ttt_i-\ttt_j} ,
\end{align}
in which the first sum is over the boundary edges of the alpha complex
incident to $x_i$,
and the second sum is over the triangles incident to these edges.
We can therefore rewrite \eqref{eqn:dd} as
\begin{align}
  d'      &=  \sum_{i,j} d_{ij} \scalprod{\uuu_{ij}}{\ttt_i-\ttt_j}
            - \sum_{i,j,k} d_{ijk} \scalprod{\uuu_{ijk}}{\ttt_i-\ttt_j} ,   \\
  d_{ij}  &=  \pi \weight{i} \frac{\sigma_{ij}}{r_i}
              \left( 1 - \frac{r_i^2-r_j^2}{\Edist{x_i}{x_j}^2} \right) , \\
  d_{ijk} &=  2 \weight{i} \frac{r_{ijk} \nu_{ijk}}
                                {r_i \Edist{x_i}{x_j}} ,
\end{align}
in which the first sum ranges over all (directed) boundary edges
of the alpha shape, and second sum ranges over all triangles incident
to these edges.
We get $\ddd_i$ by accumulating the relevant contributions.
Using $\uuu_{ji} = - \uuu_{ij}$ and $\uuu_{jik} = \uuu_{ijk}$, we get
\begin{align}
  \ddd_i  &=  \sum_j (d_{ij}+d_{ji}) \uuu_{ij}
            + \sum_{j,k} (d_{ijk}-d_{jik}) \uuu_{ijk} ,
  \label{eqn:dddi}
\end{align}
in which the first sum is over all boundary edges incident to $x_i$,
and the second sum is over all triangles incident to these edges.
Compare \eqref{eqn:dddi} with the weighted area gradient given in~\cite{BEKL04}
and with the first term of the weighted mean curvature gradient
given in \cite{AkEd19}.

\ourparagraph{Second term.}
This case is complicated by the change of motion needed to compute the
derivative of $\sigma_{ij}$.
Following the exposition in \cite[Section 4]{AkEd19}, we introduce
some notation before we can rewrite \eqref{eqn:de}.
Setting $D = \tfrac{1}{2} [1 - (r_i^2-r_j^2)/\Edist{x_i}{x_j}^2]$
and writing \mbox{$\ddd_{ijk}  =  (x_k - Dx_j + (D-1)x_i) / \Edist{x_i}{x_j}$},
we introduce three vectors so that the scalar product of the
new motion vector with $x_k - P$
can be written as a sum of three scalar products,
each involving only one of the original motion vectors:
\begin{align}
  \aaa_{ijk}  &=  x_k - P ,                                         \\
  \bbb_{ijk}  &=  [- D + \scalprod{\uuu_{ij}}{\ddd_{ijk}}] (x_k-P)
                - \scalprod{x_k-P}{\uuu_{ij}} \ddd_{ijk} ,          \\
  \ccc_{ijk}  &=  [D - 1 - \scalprod{\uuu_{ij}}{\ddd_{ijk}}] (x_k-P)
                + \scalprod{x_k-P}{\uuu_{ij}} \ddd_{ijk} ,          \\
  \scalprod{\TTT_{ijk}}{x_k-P}
    &=  \scalprod{\aaa_{ijk}}{\ttt_k}
      + \scalprod{\bbb_{ijk}}{\ttt_j}
      + \scalprod{\ccc_{ijk}}{\ttt_i} ; 
\end{align}
compare with \cite[equations (54) to (59)]{AkEd19}.
In addition, we recall the derivatives of $\sigma_{ij}$, $r_{ij}$, and $\alpha^P$ from \cite{AkEd19}:
\begin{align}
  \sigma_{ij}'  &=  \frac{1}{2 \pi r_{ij}} \left[
             \sum_k \frac{\scalprod{\TTT_{ijk}}{x_k-P}}
                         {\scalprod{x_k-P}{\uuu_{ij}^P}}
           - \sum_k \frac{\diff \Aalpha{P}}{\diff r_{ij}}
                    \frac{\diff r_{ij}}{\diff \Edist{x_i}{x_j}}
                    \scalprod{\uuu_{ij}}{\ttt_j-\ttt_i} \right] ,     \\
    \label{eqn:dsigmaij}
  \!\!\frac{\diff r_{ij}}{\diff \Edist{x_i}{x_j}}
                &=  \frac{\left( r_i^2 - r_j^2 \right)^2 - \Edist{x_i}{x_j}^4}
                    {2 \Fdist{x_i}{x_j}^2
                       \sqrt{ 2 \left( r_i^2 r_j^2 + (r_i^2+r_j^2) \Fdist{x_i}{x_j}^2 \right)
                       \!-\! \left( r_i^4 + r_j^4 + \Fdist{x_i}{x_j}^4 \right)}} , \\
  \frac{\diff \Aalpha{P}}{\diff r_{ij}}
    &=  \frac{- \scalprod{x_{ij}-P}{x_k-P}}
             {r_{ij} \sqrt{r_{ij}^2 \Edist{x_k}{P}^2 - \scalprod{x_{ij}-P}{x_k-P}^2} } ,
\end{align}
in which $\Aalpha{P}$ is the angle parametrizing the motion of the corner, $P$,
along the circle, $S_{ij}$.
With this, we rewrite \eqref{eqn:de} as
\begin{align}
  e'  &=  \sum_{i,j,k} [e_{ijk} \scalprod{\aaa_{ijk}}{\ttt_k}
                  \!+\! e_{ijk} \scalprod{\bbb_{ijk}}{\ttt_j}
                  \!+\! e_{ijk} \scalprod{\ccc_{ijk}}{\ttt_i}
                  \!+\! \bar{e}_{ijk} \scalprod{\uuu_{ij}}{\ttt_i} 
                  \!-\! \bar{e}_{ijk} \scalprod{\uuu_{ij}}{\ttt_j}],     \\
  e_{ijk}  &=  \frac{\weight{i} + \weight{j}}{4 r_{ij}}
               \frac{\lambda_{ij}}{\scalprod{x_k-P}{\uuu_{ij}^P}} ,      \\
  \bar{e}_{ijk}  &=  \frac{(\weight{i} + \weight{j}) \lambda_{ij}}{4 r_{ij}}
               \frac{\diff \alpha^P}{\diff r_{ij}}
               \frac{\diff r_{ij}}{\diff \Edist{x_i}{x_j}} ,
\end{align}
in which the sum is over the three ordered versions of all oriented
boundary triangles of the alpha shape,
and $P$ is the corresponding corner of the space-filling diagram.
Redistributing the terms, we get
\begin{align}
  \eee_i  &=  \sum_{j,k} [ e_{ijk} \aaa_{ijk}
                         + e_{kij} \bbb_{kij}
                         + e_{jki} \ccc_{jki}
                         - \bar{e}_{ijk} \uuu_{ij}
                         + \bar{e}_{jik} \uuu_{ji} ] ,
  \label{eqn:eeei}
\end{align}
in which the sum is again over the three ordered versions of
all oriented boundary triangles of the alpha shape that share $x_i$.

\ourparagraph{Third term.}
Using Lemma \ref{lem:dlambdaij}, we rewrite \eqref{eqn:df} as
\begin{align}
  f'      &=  \sum_{i,j} f_{ij} \scalprod{\uuu_{ij}}{\ttt_i-\ttt_j} ,     \\
  f_{ij}  &=  \frac{\pi}{2} (\weight{i} + \weight{j}) \sigma_{ij}
              \left[ \left( \frac{1}{2r_i} + \frac{1}{2r_j} \right)
                   - \left( \frac{1}{2r_i} - \frac{1}{2r_j} \right)
                     \frac{r_i^2-r_j^2}{\Edist{x_i}{x_j}^2} \right] ,
\end{align}
in which the sum ranges over all (directed) boundary edges of the alpha shape.
We get $\fff_i$ by accumulating the relevant contributions, as before:
\begin{align}
  \fff_i  &=  \sum_j (f_{ij} + f_{ji}) \uuu_{ij} ,
  \label{eqn:fffi}
\end{align}
in which the sum is over boundary edges incident to $x_i$.

\ourparagraph{Fourth term.}
Using Lemma \ref{lem:dphiijk}, we rewrite \eqref{eqn:dh} as
\begin{align}
  h'  &=  \sum_{i,j,k} \left[
       h_{i,jk} \scalprod{\uuu_{ij}}{\ttt_i-\ttt_j}
     + h_{j,ki} \scalprod{\uuu_{jk}}{\ttt_j-\ttt_k}
     + h_{k,ij} \scalprod{\uuu_{ki}}{\ttt_k-\ttt_i} \right] , \\
  h_{i,jk}  &=  \frac{\sigma_{ijk}}{3}
    \left[ \weight{i} p_{ij} + \weight{j} q_{ij} + \weight{k} s_{ij} \right], \\
  h_{j,ki}  &=  \frac{\sigma_{ijk}}{3}
    \left[ \weight{i} p_{jk} + \weight{j} q_{jk} + \weight{k} s_{jk} \right], \\
  h_{k,ij}  &=  \frac{\sigma_{ijk}}{3}
    \left[ \weight{i} p_{ki} + \weight{j} q_{ki} + \weight{k} s_{ki} \right],
\end{align}
in which the coefficients are defined in and after Lemma \ref{lem:dphiijk},
and the sum is over all (ordered) triangles in the boundary of the
alpha shape.
We get $\hhh_i$ by accumulating the relevant contributions:
\begin{align}
  \hhh_i  &=  \sum_{j,k}  6 (h_{i,jk} \uuu_{ij} - h_{k,ij} \uuu_{ki}) ,
  \label{eqn:hhhi}
\end{align}
in which the sum is over all ordered boundary triangles of the alpha shape
that are incident to $x_i$.
To see this, we observe that the three ordered versions of the same orientation,
$ijk, jki, kij$, all contribute the same term.
By symmetry of the spherical quadrangles defined for the opposite orientation,
the corresponding three ordered versions, $ikj, kji, jik$,
also contribute the same term.

\ourparagraph{Summary.}
We finally get the gradient of the weighted Gaussian curvature function
by adding the gradients of the four component functions:
\begin{theorem}[Gradient of Weighted Gaussian Curvature]
  \label{thm:GWGC}
  The derivative of the weighted Gaussian curvature of the space-filling
  diagram at state $\xxx$ with momentum $\ttt$ is
  \mbox{$\Diff \gauss_{\xxx} (\ttt) = \scalprod{\GGG}{\ttt}$},
  in which $\GGG_i = \ddd_i + \eee_i + \fff_i + \hhh_i$ as given
  in \eqref{eqn:dddi}, \eqref{eqn:eeei}, \eqref{eqn:fffi}, \eqref{eqn:hhhi},
  for $0 \leq i < n$.
\end{theorem}

\section{Continuity}
\label{sec:6}

Like the derivatives of the other intrinsic volumes of a
space-filling diagram, the derivative of the weighted Gaussian curvature
is almost everywhere but not everywhere continuous.
Indeed, the situation is sufficiently similar to the mean curvature case
that we can follow the analysis in \cite{AkEd19} and be brief.

\ourparagraph{General position.}
The derivative of $\gauss \colon \Rspace^{3n} \to \Rspace$ is continuous
everywhere except at non-generic states.
To define what this means, we recall that $\xxx \in \Rspace^{3n}$
encodes a collection, $X$, of $n$ closed balls in $\Rspace^3$.
We say $X$ is in \emph{general position} and, equivalently,
$\xxx$ is \emph{generic}, if the following two conditions are satisfied:
\medskip \begin{enumerate}\denselist
  \item[I.]  the common intersection of $p+1$ Voronoi domains is either empty
    or a convex polyhedron of dimension $3-p$;
  \item[II.] the common intersection of $p+1$ spheres bounding balls in $X$
    is either empty or a sphere of dimension $2-p$.
\end{enumerate} \medskip
Condition I implies that the Delaunay mosaic of $X$ is a simplicial complex
in $\Rspace^3$.
When the balls move, the mosaic undergoes combinatorial changes,
and these happen at states that violate Condition I.
In addition, Condition II captures states at which simplices in the
mosaic are added to or removed from the alpha complex.

Each violation of the two conditions corresponds to a submanifold
of dimension at most $3n-1$ in $\Rspace^{3n}$.
Write $\Mspace{I}$ and $\Mspace{II}$ for the unions of submanifolds
that correspond to violations of Condition I and of Condition II,
respectively.
Since there are only finitely many submanifolds, and at least some of
each type have dimension $3n-1$,
$\Mspace{I}$ and $\Mspace{II}$ have dimension $3n-1$ each.

\ourparagraph{Combinatorial changes.}
The submanifolds decompose $\Rspace^{3n}$ into cells,
and we refer to the connected components of
$\Rspace^{3n} \setminus (\Mspace{I} \cup \Mspace{II})$
as \emph{chambers}.
As long as the trajectory of states stays inside a chamber,
the Delaunay mosaic and the alpha complex are invariant,
combinatorially, and so are the formulas for the
weighted Gaussian curvature and its derivative.
The functions defined by the formulas are continuous,
which implies that $\gauss$ and $\nabla \gauss$ restricted
to~a~single chamber are continuous.

When the trajectory passes from one chamber to another,
it necessarily crosses a submanifold, and we let $\xxx$ be the
corresponding non-generic state.
If $\xxx \in \Mspace{I}$, then the associated combinatorial change is a flip in the
Delaunay mosaic; see \cite{Ede01} for details.
As argued in \cite{AkEd19}, $\mean$ and $\nabla \mean$
are both continuous at $\xxx$, and the same argument
implies that also $\gauss$ and $\nabla \gauss$ are continuous at $\xxx$.
If $\xxx \in \Mspace{II}$,
then the associated combinatorial change is the addition of an interval
of simplices in the Delaunay mosaic to the alpha complex or,
symmetrically, the removal of such an interval from the alpha complex.
Following \cite{AkEd19}, we distinguish between
a \emph{singular} interval, which contains only one simplex,
and a \emph{non-singular} interval, which contains two or more simplices;
see \cite{BaEd17} for background on the discrete Morse theory
of Delaunay mosaics.
A singular interval corresponds to two, three, or four spheres
whose common intersection is a single point.
If this point happens to belong to the boundary of the space-filling
diagram, then the operation changes the topology type of the surface.
By the Gauss--Bonnet theorem, this implies that in the unweighted case,
$\gauss$ changes by a~non-zero integer multiple of $2 \pi$,
and $\nabla \gauss$ is undefined.
Similarly, each non-critical interval corresponds to two, three,
or four spheres whose common intersection is a single point,
with the difference that now one of the spheres is contained
in the union of the balls bounded by the other one, two, or three spheres.
There are six basic cases, each with its own implications
on the continuity of $\nabla \gauss$.
The details are not important, so we refer to the analogous analysis
in \cite{AkEd19}.

\begin{theorem}[Continuity of Gradient]
  \label{thm:ContinuityofGradient}
  The gradient of the weighted Gaussian curvature of a space-filling diagram
  of $n$ closed balls in $\Rspace^3$ is continuous provided the state
  $\xxx \in \Rspace^{3n}$ of the diagram does not belong to $\Mspace{II}$,
  which is a $(3n-1)$-dimensional subset of $\Rspace^{3n}$.
\end{theorem}

Not all states in $\Mspace{II}$ are problematic during molecular dynamics
computations.
Some are excluded by the forces, such as two identical balls,
and others do not cause discontinuities in the derivative, such as
non-generic configurations in the interior of the space-filling diagram.
The relevant non-generic configurations are when two, three, or four
balls touch in a single point at the surface of the space-filling diagram.
For two balls, this event either merges/splits components,
or it closes/breaks a loop.
For three balls, the event either fills/opens a tunnel,
or it completes/punctures a shell surrounding a void.
For four balls, the event starts/drowns a void.

\section{Discussion}
\label{sec:7}

The main contribution of this paper is the analysis of the derivative
of the weighted Gaussian curvature of a space-filling diagram.
Specifically, we give an explicit description of the gradient of
the weighted Gaussian curvature function, which for $n$ spheres is
a map from $\Rspace^{3n}$ to $\Rspace$.
In addition, we study the subset of $\Rspace^{3n}$ at which
the derivative violates continuity.
In summary, this is sufficient information for an efficient and robust
implementation of~the~weighted Gaussian curvature derivative,
one that can be added to the inner loop of a~molecular dynamics simulation
of a physical system.

\subsection*{Acknowledgment}
\footnotesize{
The authors of this paper thank Roland Roth for suggesting the analysis of
the weighted curvature derivatives for the purpose of improving molecular
dynamics simulations.
They also thank Patrice Koehl for the implementation of~the formulas
and for his encouragement and advise along the road.
Finally, they thank two anonymous reviewers for their constructive criticism.
}

\bigskip 

\noindent


\appendix \clearpage
\normalsize
\section{Proof of Formula \ref{form:ProductofSines}}
\label{app:A}

We present a proof of the formula from Section \ref{sec:4}
after restating it for convenience.

\bigskip \noindent
\textbf{Formula 3} ({\sf Product of Sines}).
  \textit{Writing $a = \cos^2 \frac{\phi_{ij}}{2}$,
  $b = \cos^2 \frac{\phi_{jk}}{2}$, and $c = \cos^2 \frac{\phi_{ki}}{2}$,
  the expression under the square root in \eqref{eqn:sphericalarea} satisfies
  \begin{align}
    \sin s \sin (s-\phi_{ij}) \sin (s-\phi_{jk}) \sin (s-\phi_{ki})
      &=  4abc - (a+b+c-1)^2 .
  \end{align}}
\vspace{-0.2in}
\ourproof
  Recall the trigonometric identity
  $\sin (\alpha + \beta + \gamma)
      = \cos \alpha \cos \beta \sin \gamma + \cos \alpha \sin \beta \cos \gamma
      + \sin \alpha \cos \beta \cos \gamma - \sin \alpha \sin \beta \sin \gamma$.
  Setting
  \begin{align}
    A  &=  \cos \tfrac{\phi_{ij}}{2} \cos \tfrac{\phi_{jk}}{2} \sin \tfrac{\phi_{ki}}{2}, ~~~~
    B   =  \cos \tfrac{\phi_{ij}}{2} \sin \tfrac{\phi_{jk}}{2} \cos \tfrac{\phi_{ki}}{2}, \\
    C  &=  \sin \tfrac{\phi_{ij}}{2} \cos \tfrac{\phi_{jk}}{2} \cos \tfrac{\phi_{ki}}{2}, ~~~~
    D   =  \sin \tfrac{\phi_{ij}}{2} \sin \tfrac{\phi_{jk}}{2} \sin \tfrac{\phi_{ki}}{2},
  \end{align}
  and observing that $\cos (- \alpha) = \cos \alpha$
  as well as $\sin (- \alpha) = - \sin \alpha$,
  we get
  \begin{align}
    \sin s                &=  \sin \left( \tfrac{\phi_{ij}}{2} + \tfrac{\phi_{jk}}{2}
                                                               + \tfrac{\phi_{ki}}{2} \right)
                           =  A + B + C - D , \\
    \sin (s - \phi_{ij})  &=  \sin \left( - \tfrac{\phi_{ij}}{2} + \tfrac{\phi_{jk}}{2}
                                                               + \tfrac{\phi_{ki}}{2} \right)
                           =  A + B - C + D , \\
    \sin (s - \phi_{jk})  &=  \sin \left( \tfrac{\phi_{ij}}{2} - \tfrac{\phi_{jk}}{2}
                                                               + \tfrac{\phi_{ki}}{2} \right)
                           =  A - B + C + D , \\
    \sin (s - \phi_{ki})  &=  \sin \left( \tfrac{\phi_{ij}}{2} + \tfrac{\phi_{jk}}{2}
                                                               - \tfrac{\phi_{ki}}{2} \right)
                           =  - A + B + C + D .
  \end{align}
  Multiplying the four right-hand sides, we get
  $-(A^4 + B^4 + C^4 + D^4) + 8ABCD
   + 2(A^2B^2 + A^2C^2 + A^2D^2 + B^2C^2 + B^2D^2 + C^2D^2)$.
  To rewrite this expression, we recall that
  \begin{align}
    A^4    &=  [a b (1-c)]^2 ,                             ~~~~~~~~~~~~~~~
    B^4     =  [a (1-b) c]^2 ,                                          \\
    C^4    &=  [(1-a) b c]^2 ,                             ~~~~~~~~~~~~~~~      
    D^4     =  [1 - a - b - c + ab + ac + bc - abc]^2 ,                 \\
    ABCD   &=  a (1-a) b (1-b) c (1-c) ,                                \\
    A^2B^2 &=  a^2 b (1-b) c (1-c) ,                                   ~~~
    A^2D^2  =  a (1-a) b (1-b) c (1-c)^2 ,                              \\
    A^2C^2 &=  a (1-a) b^2 c (1-c) ,                                   ~~~
    B^2D^2  =  a (1-a) (1-b)^2 c (1-c) ,                                \\
    B^2C^2 &=  a (1-a) b (1-b) c^2 ,                                   ~~~
    C^2D^2  =  (1-a)^2 b (1-b) c (1-c) .                              
  \end{align}
  Developing all the right-hand sides and canceling the terms of degree
  4, 5, 6 and more, we get the claimed relation.
\eop

\newpage
\section{Notation}
\label{app:N}

\begin{table}[hbt]
  \centering \begin{tabular}{ll}
    $S_i = \boundary{B_i}$
      &  sphere bounding ball                                           \\
    $S_{ij} = \boundary{B_{ij}}$
      &  circle bounding disk                                           \\
    $S_{ijk} = \boundary{B_{ijk}}$
      &  pair of points bounding line segment                           \\
    $x_i, x_{ij}, x_{ijk}$
      &  centers of $S_i$, $S_{ij}$, $S_{ijk}$                          \\
    $r_i, r_{ij}, r_{ijk}$
      &  radii of $S_i$, $S_{ij}$, $S_{ijk}$                            \\
    $\uuu_{ij} = \frac{x_i - x_j}{\Edist{x_i}{x_j}}$
      &  unit vector between centers                                    \\
    $\uuu_{ijk} = \tfrac{\uuu_{ik} - \scalprod{\uuu_{ik}}{\uuu_{ij}} \uuu_{ij}}
                  {\|{\uuu_{ik} - \scalprod{\uuu_{ik}}{\uuu_{ij}} \uuu_{ij}}\|}$
      &  unit normal to $\uuu_{ij}$ with positive component in direction $\uuu_{ik}$  \\
    $\nu_i = \frac{\Volume(B_i \cap \Vdom{i})}{\Volume(B_i)}$
      &  volume fraction of ball                                        \\
    $\nu_{ij} = \frac{\Area(B_{ij} \cap \Vdom{ij})}{\Area(B_{ij})}$
      &  area fraction of disk                                          \\
    $\nu_{ijk} = \frac{\Length(B_{ijk} \cap \Vdom{ijk})}{\Length(B_{ijk})}$
      &  length fraction of line segment                                \\
    $\nu_{ijk\ell} = \frac{\Card{B_{ijk\ell} \cap \Vdom{ijk\ell}}}
                          {\Card{B_{ijk\ell}}}$
      &  $0$ or $1$                                                     \\
    $\sigma_i = \frac{\Area(S_i \cap \Vdom{i})}{\Area(B_i)}$
      &  area fraction of sphere                                        \\
    $\sigma_{ij} = \frac{\Length(S_{ij} \cap \Vdom{ij})}{\Length(S_{ij})}$
      &  length fraction of circle                                      \\
    $\sigma_{ijk} = \frac{\Card{S_{ijk} \cap \Vdom{ijk}}}{\Card{S_{ijk}}}$
      &  $0$, $\tfrac{1}{2}$, or $1$                                    \\
    $\weight{i}; \phi_{ij}; \lambda_{ij}$
      &  real weight; angle between normals; combined length of projection  \\
    $\phi_{i,jk} = \alpha_i \, \phi_{ijk}$
      &  area of spherical quadrangle; area fraction, area of spherical triangle \\
                                                                  \\
    $X, \bigcup X$
      &  set of balls, space-filling diagram                      \\
    $\Fspace, \Sspace^2, \Bspace^3, \Rspace^3$
      &  surface, unit sphere, unit ball, Euclidean space         \\
    $\xxx, \ttt; \vvv, \aaa, \mmm, \GGG$
      &  state, momentum; gradients                               \\
    $F \colon \Rspace^{3n} \to \Rspace$
      &  intrinsic volume function                                \\
    $\kappa_1(x), \kappa_2(x); \Euler{\Fspace}$
      &  principal curvatures; Euler characteristic               \\
    $P, \alpha^P; \xi_i$
      &  intersection point, parametrizing angle; signed distance \\
    $\nnn_i, \mmm_i, \zzz; \alpha_i$
      &  unit normal, midpoint, center; area fraction             \\
    $a, b, c; s$
      &  squared cosines; half-perimeter                          \\
    $S, A, B, C$
      &  area functions                                           \\
    $\sgm{i,jk}$
      &  orientation of isosceles triangle                        \\
    $R_{ijk}; r$
      &  radius; squared cosine of half-radius                    \\
    $u, v, w, t$
      &  generic functions                                        \\
    $\gauss' = d' + e' + f' + h'$
      &  derivatives                                              \\
    $\ddd, \eee, \fff, \hhh; \ddd_i, \eee_i, \fff_i, \hhh_i$
      &  gradients; as they apply to $x_i$                        \\
    $p_{ij}, q_{ij}, s_{ij}$
      &  coefficients                                             \\
    $d_{ij}, d_{ijk}, e_{ijk}, \bar{e}_{ijk}, f_{ij}, h_{i,jk}$
      &  coefficients                                             \\
    $D; \aaa_{ijk}, \bbb_{ijk}, \ccc_{ijk}, \ddd_{ijk}$
      &  auxiliary constant; auxiliary vectors                    \\
    $\Mspace{I}, \Mspace{II}$
      &  subsets of $\Rspace^{3n}$                                \\
  \end{tabular}
  \caption{Notation for concepts, sets, functions, vectors, variables.}
  \label{tbl:Notation}
\end{table}

\end{document}